\documentclass[conference]{IEEEtran}
\IEEEoverridecommandlockouts

\usepackage{lipsum}
\usepackage{cite}
\usepackage{amsmath,amssymb,amsfonts, amsthm}
\usepackage{algorithmic, algorithm}
\usepackage{graphicx}
\usepackage{textcomp}
\usepackage{xcolor}
\usepackage{verbatim}

\usepackage{booktabs}
\usepackage{multirow}
\usepackage{enumitem, array}
\usepackage{caption}
\usepackage{subcaption}
\usepackage[export]{adjustbox}
\usepackage{hyperref}
\usepackage{gensymb}

\usepackage{colortbl}

\usepackage{multirow}

\def\BibTeX{{\rm B\kern-.05em{\sc i\kern-.025em b}\kern-.08em
    T\kern-.1667em\lower.7ex\hbox{E}\kern-.125emX}}

\def\T{{\bf T}}

\def\Y{{\bf Y}}
\def\t{{\bf t}}
\def\x{{\bf x}}
\def\f{{\bf f}}
\def\y{{\bf y}}

\def\w{{\bf \omega}}

\newtheorem{thm}{Theorem}[section]

\newtheorem{prop}[thm]{Proposition}

\newtheorem{cor}{Corollary}

\theoremstyle{definition}

\newcommand{\Var}{\mathrm{Var}}

\newcommand{\update}[1]{{\color{black} #1}}

\begin{document}

\title{Provable Performance Guarantees of Copy Detection Patterns
\thanks{S. Voloshynovskiy is a corresponding author.}
\thanks{This research was partially funded by the Swiss National Science Foundation SNF No. 200021\_182063.}}

\author{
\IEEEauthorblockN{Joakim Tutt}
\IEEEauthorblockA{\textit{Department of Computer Science} \\
\textit{University of Geneva}\\
Geneva, Switzerland \\
joakim.tutt@unige.ch}
\and
\IEEEauthorblockN{Slava Voloshynovskiy}
\IEEEauthorblockA{\textit{Department of Computer Science} \\
\textit{University of Geneva}\\
Geneva, Switzerland \\
svolos@unige.ch}
}

\maketitle

\begin{abstract}
Copy Detection Patterns (CDPs) are crucial elements in modern security applications, playing a vital role in safeguarding industries such as food, pharmaceuticals, and cosmetics. Current performance evaluations of CDPs predominantly rely on empirical setups using simplistic metrics like Hamming distances or Pearson correlation. These methods are often inadequate due to their sensitivity to distortions, degradation, and their limitations to stationary statistics of printing and imaging. Additionally, machine learning-based approaches suffer from distribution biases and fail to generalize to unseen counterfeit samples. Given the critical importance of CDPs in preventing counterfeiting, including the counterfeit vaccines issue highlighted during the COVID-19 pandemic, there is an urgent need for provable performance guarantees across various criteria. This paper aims to establish a theoretical framework to derive optimal criteria for the analysis, optimization, and future development of CDP authentication technologies, ensuring their reliability and effectiveness in diverse security scenarios.
\end{abstract}


\section{Introduction}
\label{sec:introduction}

Copy Detection Patterns (CDPs) have emerged as indispensable tools in modern security frameworks, particularly in sectors where the integrity and authenticity of products are paramount \cite{picard2004digital,picard2021counterfeit,tkachenko:2021review}. Industries such as food, pharmaceuticals, and cosmetics heavily rely on CDPs to combat counterfeiting, ensuring consumer safety and maintaining product integrity. The urgency of this issue has been underscored by the proliferation of counterfeit vaccines during the COVID-19 pandemic, highlighting the critical need for reliable anti-counterfeiting measures.
CDPs are essentially random binary patterns that possess high entropy, meaning the probability of black and white dots in these codes is equal to $0.5$. These patterns are reproduced on physical media using high-resolution printing or laser engraving. Authentication of CDPs involves comparing the physical patterns to their corresponding digital templates. The fundamental idea is that any attempt to counterfeit these patterns will introduce additional distortions during scanning and reproducing processes, allowing the authentication system to distinguish between genuine and fake CDPs.

The challenge in CDP authentication is twofold. Firstly, attackers might use machine learning techniques to produce high-quality fakes. They can scan the printed CDPs from physical objects, preprocess them to estimate the original digital templates, and then reprint these estimates. These attacks, known as machine learning-based copy attacks, represent the most sophisticated and challenging threat, as described in \cite{chaban2021machine, yadav2019estimation,khermaza2021can,taran2019clonability}. Secondly, the imaging device used to acquire the patterns introduces additional degradation during the acquisition process, further complicating the authentication task.
Another significant technical constraint is the need to authenticate CDPs based solely on the digital templates from which they are printed. Authentication based on physically scanned patterns is impractical due to the difficulty of managing the massive production volumes. Therefore, this study focuses on technologies that can authenticate printed CDPs using modern digital printing techniques and mobile phone acquisitions, relying only on the digital templates.
Furthermore, the current methods for authenticating  CDPs are often inadequate. Empirical setups typically utilize simplistic metrics like Hamming distances or Pearson correlation. While these metrics offer some insights, they fall short when accounting for the distortions and degradations that CDPs may encounter in real-world scenarios \cite{Chaban2022wifs}.
Machine learning methods, on the other hand, present their own set of challenges. Although they can provide sophisticated analysis, these methods are prone to distribution biases. When faced with new counterfeit samples that were not part of the training data, these models often fail to perform reliably \cite{pulfer2022icip}. This limitation is especially concerning given the dynamic and evolving nature of counterfeiting tactics.

\begin{figure*}[t]
\centering
{\includegraphics[scale=.23, clip]{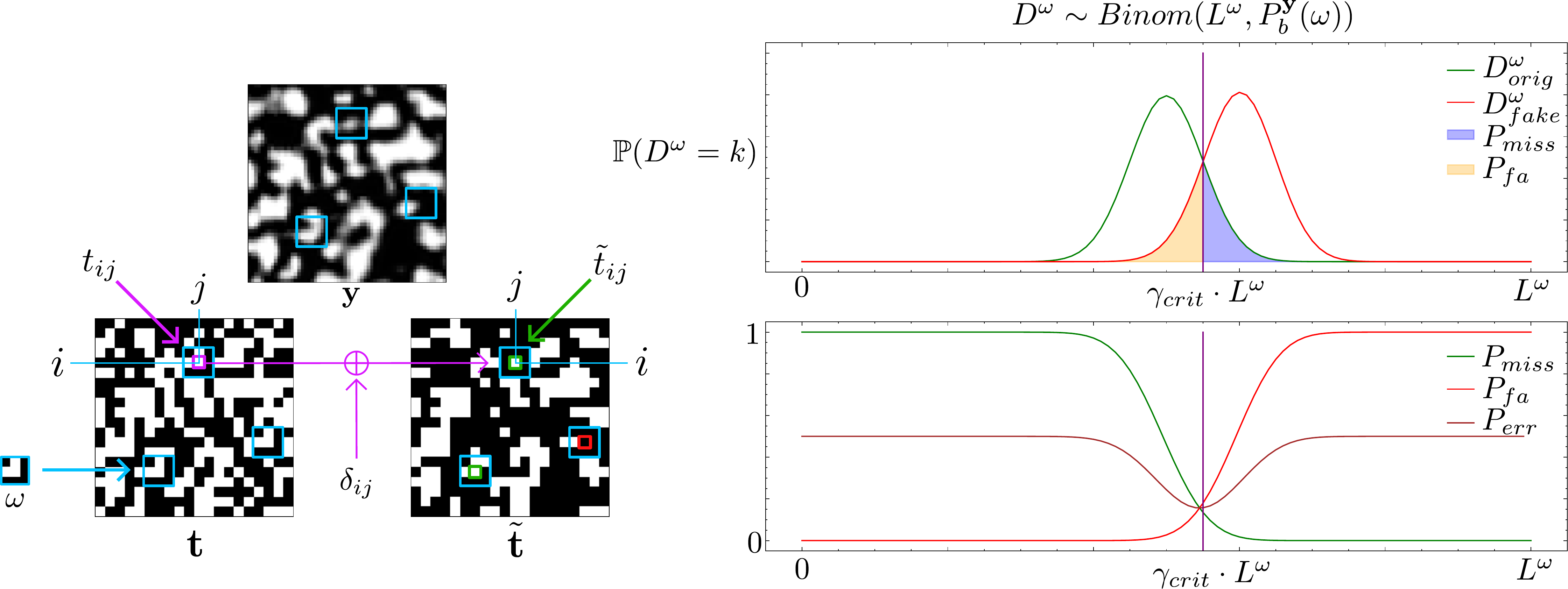}}

\caption{An illustration of the theoretical framework under study. Given a fixed pattern $\w$ and a CDP probe $\y$, one identifies all appearances of the pattern $\w$ (left) in the digital template $\t$. For each location $(i,j)$ of $\w$ in $\t$, one can associate a random bit-flipping $\delta_{i,j}$. The Hamming distance $D^{\w}$ between $\t$ and $\tilde{\t}$ for the pattern $\w$ follows a Binomial distribution which is compared with the reference Binomial distributions of originals and fakes CDP. Based on the reference distributions, an optimal separation bound $\gamma_{crit}$ is computed which minimizes the average probability of error.}
\label{fig:full_setup_authentication}
\end{figure*}

In light of these challenges, there is a pressing need to develop a comprehensive analytical framework that provides provable guarantees on the performance of CDPs across a range of criteria. Such a framework would not only enhance the reliability and effectiveness of CDPs but also guide future developments in authentication technologies. This paper seeks to address this need by proposing a theoretical approach to derive optimal criteria for the analysis and optimization of CDP performance. Through this framework, we aim to ensure that CDP technologies can be systematically improved to meet the rigorous demands of modern security applications.

In previous publications \cite{tutt2022wifs,tutt2024pattern}, we demonstrated that different CDP patterns, composed of $3 \times 3$ elements, have varying probabilities of bit-flipping after printing, scanning or mobile phone image acquisition and binarization. We identified 512 different patterns and characterized their reliability based on the context of their $3 \times 3$ neighborhoods. However, those studies employed simple rules, using masking based on empirically optimized thresholds, and focused only on reliable channels characterized by low bit-flipping probabilities. Patterns with high bit-flipping probabilities were disregarded completely. Furthermore, simple metrics such as Hamming distance or Pearson correlation were used for the reliable patterns based CDP authentication.

In contrast, this new study explores more sophisticated rules that go beyond hard decision-making and the binary characterization of reliable and unreliable patterns. Instead, this paper introduces three different criteria for the measure of quality of the acquired CDP based on well-known statistical tests namely: The Hamming distance test; the cross-entropy between the probability of bit-flipping and a reference probability; and Neymann-Pearson likelihood ratio.
We investigate different fusion rules and decision strategies for binarized CDP. Specifically, we consider two approaches: making a decision first and then aggregating the results versus aggregating everything first and then making a decision. This paper presents theoretical findings on how to optimize these strategies for enhanced CDP authentication.

\section{Information flow in Copy Detection Patterns}

\label{sec:info_flow_in_cdp}

In this section, we reformulate the classical authentication process of CDP using an information theoretic approach \cite{Beekhof:2012ys}. When trying to identify a probe sample $\y$, one typically uses binarization and compares the binarized probe $\tilde{\t}$ with the original template $\t$ using Hamming distance $d(\t,\tilde{\t})$. A decision threshold $\gamma \in [0,1]$ is then fixed to decide whether the probe $\y$ is genuine or counterfeited:

\begin{equation}
d(\t, \tilde{\t}) \gtrless \gamma B,
\end{equation}
where $B$ denotes the total number of symbols in $\t$.
From the point of view of information theory and following the analysis done in \cite{tutt2024pattern}, this process can be described as a Markov chain:
$\T \rightarrow \Y \rightarrow \tilde{\T},$
where $\T$ is a random binary matrix in which each symbol has equal probability of being black or white, $\Y$ is a random matrix with real value entries between $0$ and $1$ and $\tilde{\T}$ is a binary matrix with fixed probability for each symbol.
In this framework, each symbol transition can be described as a noisy binary channel:
\begin{equation}
\label{eq:binary_channel_ij}
\tilde{T}_{ij} = T_{ij} \oplus \delta_{ij},
\end{equation}
where $P_b(i,j) = \mathbb{P}(T_{ij} \neq \tilde{T}_{ij})$ is the probability of bit-flip for symbol $T_{ij}$, $\delta_{ij} = 1$ if there is a bit-flip and $\delta_{ij} = 0$, otherwise (see Fig. \ref{fig:full_setup_authentication}, left). In this setting, the focus is on being able to describe the posterior distribution $\mathbb{P}(\tilde{\T} | \T)$.

The main contribution of \cite{tutt2024pattern} was to precisely describe the variability of the probability of bit-flipping $P_b(i,j)$ as a function of the pixel position $(i,j)$ by introducing \update{the Binary Pattern-based Channel (BPC) model}.
This model makes the assumption that the probability of bit-flipping of the symbol $T_{ij}$ only depends on a small subset of symbols directly surrounding symbol $T_{ij}$, which we denote as $\w_{ij}$. Typically:
\begin{equation}
\omega_{ij} = \{ T_{i \pm a, j \pm b} | 0 \leq a,b < h/2 \},
\end{equation}
for an $h \times h$ square pattern centered around $T_{ij}$. Apart from this local dependency, the model assumes independence of the probability of bit-flipping with respect to the location of the pattern. That is, if $\w_{ij} = \w_{kl}$ are two identical patterns located in different coordinates, then:
\begin{equation}
P_b(i,j) = P_b(k,l).
\end{equation}
Based on these assumptions, the characterization of the binary channels $T_{ij} \rightarrow \tilde{T}_{ij}$ reduces to the study of the binary channels $T_{\w} \rightarrow \tilde{T}_{\w}$, where $T_{\w}$ indicates the central symbol of pattern $\w$. The advantage of this approach is the reduction of dimension, from the study of $B$ different channels to $M = 2^{h^2}$ (in our experiments,  $B = 51'984$, $h = 3$ and $M = 512$).
According to this model, one can compute the Hamming distance for each pattern $\w$:
\begin{equation}
D^{\w} = d^{\w}(\T, \tilde{\T}) = \sum\limits_{\substack{i,j: \\ \w_{ij} = \w}} \delta_{ij},    
\end{equation}
and compute a pattern-wise probability of bit-flipping:
\begin{equation}
P^{\y}_b(\w) = \frac{1}{L^\w} \mathbb{E}[{D^\w}],
\end{equation}
where $L^\w$ denotes the number of pattern $\w$ appearing in $\T$ and $P^{\y}_b(\w)$ is the probability of bit-flipping for pattern $\w$. As such, the model transforms a digital probe $\y$  into a feature vector of size $M$: $\{P_b^\y(\w)\}_{\w = 0}^{M-1}.$\footnote{Profitting from the fact that, on average, $L^{\w} = B / M \approx 100$, one can estimate the probability of bit-flipping empirically for each of the patterns solely based on one pair $(\t,\tilde{\t})$ and build a feature vector
for a given probe~$\y$.}

\section{A statistical viewpoint}

\label{sec:statistical_viewpoint}

In this paper, we use the BPC model \cite{tutt2024pattern} as a baseline and describe the problem of authentication of CDP as a collection of independent channel-wise statistical tests. Under this model, each pattern $\w$ gives rise to a random binomial variable $D^{\w}$ independent of each other. Based on this mathematical description, we focus on different ways to measure channel performances and study how they can be optimally combined together to strengthen authentication. That is, given a probe $\y$, decide whether it is an original $\x$ or a fake $\f$. We denote $\mathbb{H}_{0}$ the hypothesis that probe $\y$ is an original and $\mathbb{H}_{1}$ the hypothesis that $\y$ is a fake.
Under $\mathbb{H}_{0}$, the reference original codebook is denoted as $\mathcal{C}_{0} = \{ P_b(\w) \}_{\w=0}^{M - 1}$ while, under $\mathbb{H}_{1}$, the reference fake codebook is denoted as $\mathcal{C}_{1} = \{ Q_b(\w) \}_{\w = 0}^{M - 1}$.
Given an unidentified probe $\y$, and a fixed pattern $\w$, we perform the following steps (see Fig. \ref{fig:full_setup_authentication} for an illustration):

\begin{itemize}
    \item Identify each appearance of pattern $\w$ in the reference template $\t$ and extract the central pixel of each of these patterns: $t_1, ..., t_{L^\w}$;
    \item Binarize the probe $\y \mapsto \tilde{\t}$ and extract the corresponding central pixels: $\tilde{t}_1, ..., \tilde{t}_{L^\w}$;
    \item Compute the bit-flipping values $\delta_k=t_k \oplus \tilde{t}_k$ for each $k = 1,...,L^\w$, where $\delta_k$ denotes the indicator function of bit-flipping with $\delta_k = 1$, if $t_k  \neq \tilde{t}_k $, and 0, otherwise. 
\end{itemize}
The BPC model defines that the bit-flipping values $\delta_1, ..., \delta_{L^\w}$ are sampled according to a Bernoulli distribution with a fixed probability $P^{\y}_b(\w)$ which can be estimated empirically as:
\begin{equation}
\label{eq:P_b_approximation_formula}
P^{\y}_b(\w) \approx \frac{1}{L^\w} \sum_{k=1}^{L^\w} \delta_{k}.
\end{equation}
The goal of authentication is now to decide whether this empirical probability is more likely to be $P_b(\w)$ or $Q_b(\w)$ and provide bounds on this decision \cite{cover2012chapter11}.

\subsection{Hamming distance test}
\label{ssec:hamming_distance_test}

In the following sections, all analyses are conducted on a fixed pattern $\w$ and then extended to all $\w$. As such, and to ease the notations, we remove the explicit reference to the pattern in the indices, except when formulating final results:
$$D^\w \rightarrow D, \quad L^\w \rightarrow L, \quad P_b(\w) \rightarrow P_b, \quad Q_b(\w) \rightarrow Q_b.$$
Following the analysis in \cite{Beekhof:2012ys}, the Hamming distance is described by a binomial distribution:
\begin{equation}
\label{eq:def_D_hamm_dist}
D = \sum_{k=1}^{L} \delta_k \sim Binom(L ; P^{\y}_b).
\end{equation}
If we suppose that we have access to reference codebooks for original and fakes, we can formulate the Hamming distance test and measure the probability of miss as:
\begin{IEEEeqnarray}{lCl}
P_m(\gamma) & = & \mathbb{P}_{\mathbb{H}_{0}}(D > \gamma L) \nonumber \\
& = & \sum_{k=\lfloor \gamma L \rfloor+1}^{L} \binom{L}{k} P_b^k (1-P_b)^{L-k},    
\end{IEEEeqnarray}
and the probability of \update{false acceptance} as:
\begin{equation}
P_{fa}(\gamma) = \mathbb{P}_{\mathbb{H}_{1}} (D \leq \gamma L)  = \sum_{k=0}^{\lfloor \gamma L\rfloor} \binom{L}{k} Q_b^k (1-Q_b)^{L-k}.
\end{equation}
One way to choose an optimal decision threshold $\gamma$ is to minimize the average probability of error as:

\begin{equation}
\min_{\gamma \in [0,1]} P_{err}(\gamma) = \min_{\gamma \in [0,1]} \frac{P_m(\gamma) + P_{fa}(\gamma)}{2},
\end{equation}
assuming both hypotheses $\mathbb{H}_0$ and $\mathbb{H}_1$ are equiprobable (see Fig. \ref{fig:full_setup_authentication}, \update{right}).
Our goal, as defined in the following sections, is to better study these probability of errors and derive an optimal bound for each channel by making an explicit link between this approach and the Neymann-Pearson test.

\subsection{Posterior likelihood approach}

\label{sec:posterior_likelihood}

The authentication problem can be studied from the point of view of the likelihood function. Given a sequence of realisations $\delta_1, ..., \delta_L \in \{0, 1\}$ and a reference probability of bit-flipping $P_b$, one can measure the likelihood that these samples were drawn from a Bernoulli distribution with parameter $P_b$. We write:
\begin{IEEEeqnarray}{lCl}
\label{eq:link_likelihood_hamming_dist}
\mathbb{P} (\tilde{t}_1, ..., \tilde{t}_{L} | t_1, ..., t_L ) & = & \prod_{k=1}^L  \mathbb{P}({\tilde{t}}_k| {t}_k) \nonumber \\
& = & \prod_{k=1}^{L} P_b^{\delta_k}  \cdot (1 - P_b)^{1 - \delta_k} \nonumber \\
& = & P_b^{D} \cdot (1 - P_b)^{L - D},
\end{IEEEeqnarray}
where $D$ is the Hamming distance defined in \eqref{eq:def_D_hamm_dist} and follows a Binomial law. Note that this derivation is based on the assumption of the BPC model that each bit-flipping $\delta_k$ is identically distributed and independent. Taking the logarithm on both sides, one arrives to the formula:
\begin{equation}
\label{eq:link_log_likelihood_hamming_dist}
\log \mathbb{P} (\tilde{t}_1, ..., \tilde{t}_{L} | t_1, ..., t_L )= D \cdot \log P_b+ (L-D) \cdot \log \left(1-P_b\right).
\end{equation}
These types of derivations are well known \cite{cover2012chapter11} and explicit the link between minimization of Hamming distance and maximization of log-likelihood. Taking the expectation with respect to the posterior distribution on both sides yields the following result:

\begin{prop}
\label{thm:avg_log_likelihood_cross_entropy}
The average log-likelihood is given by:
\begin{equation}
\label{eq:likelihood_and_cross_entropy}
\frac{1}{L^\w} \mathbb{E}[ \log \mathbb{P} (\tilde{t}_1, ..., \tilde{t}_{L^\w} | t_1, ..., t_{L^\w} ) ] = - \mathcal{H}(P_b^{\y}(\w) ; P_b(\w)),    
\end{equation}
where the right-hand side designates the cross-entropy between the unknown probe distribution $P_b^{\y}(\w)$ and the reference distribution $P_b(\w)$.
\end{prop}

\begin{proof}
    The proof is straightforward from \eqref{eq:link_log_likelihood_hamming_dist} since $D~\sim~Binom(L, P_b^{\y})$, and thus $\frac{1}{L} \mathbb{E}[D] = P_b^{\y}$.
\end{proof}

Formula \eqref{eq:likelihood_and_cross_entropy} is very useful to test the adequacy of probe $\y$ to a certain reference probability $P_b$. In practice one can use \eqref{eq:P_b_approximation_formula} to estimate these quantities empirically.
This result also gives new insights on the Posterior Log Likelihood (PLL) metric introduced in \cite{tutt2022wifs}.

\begin{cor}
The PLL metric can be computed as an average on all channels:
\begin{equation}
    PLL(\y ; \mathcal{C}) = - \sum_{\w = 0}^{M - 1} L^\w \cdot \mathcal{H}(P_b^{\y}(\w) ; P_b(\w)).
\end{equation}
\end{cor}

\subsection{Neymann-Pearson optimal bound}

Using Neymann-Pearson terminology, we will designate the type I error in this problem, as accepting the probe as an original when it is a fake (false acceptance). The type II error would be to reject a probe when it is an original (true miss). The problem is, by nature, asymmetrical as it is much more harmful in real life scenarios to accept a fake product than to throw away a genuine one (think for instance of fake drugs).
Neymann-Pearson lemma \cite{Neyman:1933wgr}, then provides an optimal decision criterion for a significance test at level $\alpha$. That is, if we fix the Type I error to be equal to $\alpha$, the optimal decision statistic is given by the likelihood ratio test:
\begin{equation}
\label{eq:NP_ratio_test}
\mathcal{R} = \frac{\mathbb{P}_{\mathbb{H}_{0}}(\tilde{t}_1, ..., \tilde{t}_L| t_1, ..., t_L)}{\mathbb{P}_{\mathbb{H}_{1}}(\tilde{t}_1, ..., \tilde{t}_L| t_1, ..., t_L)} > \rho.
\end{equation}

\begin{prop}
    \label{thm:likelihood_test_asymptotic}
    The likelihood ratio test \eqref{eq:NP_ratio_test} is equivalent to:
    \begin{equation}
    \label{eq:NP_CE_diff_test}
    \mathcal{H}(P_b^\y(\w) ; Q_b(\w)) - \mathcal{H}(P_b^\y(\w) ; P_b(\w)) > \frac{\log \rho}{L^{\w}},
    \end{equation}
    and this test is asymptotically optimal (with $L^{\w} \rightarrow +\infty$) as long as $Q_b(\w) \neq P_b(\w)$.
\end{prop}

\begin{proof}
    By computing $\frac{1}{L} \log \mathcal{R}$ and following the same derivation as \eqref{eq:link_likelihood_hamming_dist}, one arrives to the following formula:
    \begin{equation}
    \label{eq:log_ratio_and_Hamming_link}
    \frac{1}{L} \log \mathcal{R} = \frac{D}{L} \log \frac{P_b}{Q_b} + \frac{L-D}{L} \log \frac{1 - P_b}{1 - Q_b}.
    \end{equation}
    Equation \eqref{eq:NP_CE_diff_test} then follows by taking the expectation on both sides as in  \eqref{eq:likelihood_and_cross_entropy}. 
    Under hypothesis $\mathbb{H}_{0}$, the random variable $D~\sim~Binom(L ; P_b)$ follows a binomial distribution and thus:
    \begin{IEEEeqnarray}{lCl}
    \frac{1}{L} \mathbb{E}[ \log \mathcal{R} | \mathbb{H}_{0}] & = & P_b \cdot \log \frac{P_b}{Q_b} + (1-P_b) \cdot \log \frac{1 - P_b}{1 - Q_b} \nonumber \\
    & = & \mathcal{D}_{KL}(P_b || Q_b),
    \end{IEEEeqnarray}
    where $\mathcal{D}_{KL}(P_b || Q_b)$ denotes the Kullback-Leibler divergence between the Bernoulli distributions given by $P_b$ and $Q_b$.
    With similar computations one can show that:
    \begin{equation}
    \frac{1}{L} \mathbb{E}[ \log \mathcal{R} | \mathbb{H}_{1}] = - \mathcal{D}_{KL}(Q_b || P_b).
    \end{equation}
    Finally, with similar approach, one shows that the variance $\Var(\frac{\log \mathcal{R}}{L}) \rightarrow 0$, when $L \rightarrow + \infty$ and so these quantities concentrate around their means, yielding asymptotic optimality, independant of $\rho$.    
\end{proof}

Note that \eqref{eq:NP_CE_diff_test} can be seen as a direct extension of the approach in \cite{tutt2024pattern}. Indeed, the PLL metric was solely based on a single cross-entropy term, whereas the Neymann-Pearson test is based on the difference of two cross-entropies. PLL can thus be considered as an analogy to a one-class decision based solely on $P_b$, while NP-test is a two-class decision which requires the knowledge of both $P_b$ and $Q_b$.

\begin{prop}
    The Neymann-Pearson test and the Hamming distance test $\mathcal{R} > \rho$ and $D < \gamma L$
    are equivalent if $P_b \leq Q_b$ and:
    \begin{equation}
    \label{eq:log_rho_equals_diff_D_KL}
        \frac{\log \rho}{L} = \mathcal{D}_{KL}(\gamma || Q_b) - \mathcal{D}_{KL}(\gamma || P_b).
    \end{equation}
    When $\rho = 1$, that is when both likelihoods are equal, we can derive the critical separation boundary independent of $L$:
    \begin{equation}
    \label{eq:def_gamma_crit}
        \gamma_{crit} = \left( {1 + \frac{\log P_b  - \log Q_b}{\log (1 - Q_b) - \log (1 - P_b)}}\right)^{-1}.
    \end{equation}
    
\end{prop}

\begin{proof}
    To prove \eqref{eq:log_rho_equals_diff_D_KL}, we start with  relation \eqref{eq:log_ratio_and_Hamming_link} and use the facts that $D < \gamma L$, $\log(\frac{P_b}{Q_b}) \leq 0$ and $\log(\frac{1 - P_b}{1 - Q_b}) \geq 0$ to derive the lowerbound \eqref{eq:log_rho_equals_diff_D_KL}.
    Equation \eqref{eq:def_gamma_crit} is obtained from \eqref{eq:log_rho_equals_diff_D_KL} by setting $\rho = 0$ and isolating $\gamma$.
\end{proof}

\section{Channels aggregation strategies}
\label{sec:aggregation_strategies}

All statistical approaches described in Section \ref{sec:statistical_viewpoint}, were computed independently for each channel and different approaches to optimal thresholds selection were given. Applying a decision threshold to each channel transforms the real-valued feature vector $\{ P_b^{\y}(\w) \}_{\omega=0}^{M-1}$ into a binary decision vector: $\{ \Delta_{\w} \}_{\w=0}^{M-1}$, where $\Delta_\w \in \{ 0 , 1 \}$.
In both situations, we are left with the problem of optimally aggregating these pattern-wise scores into a final decision score $S_{final}$. We propose here a generic way of describing linear aggregation strategies.
A linear aggregation strategy is a weighted sum:
\begin{equation}
    S_{final}^{AD} = \sum_{\w} P_b^{\y}(\w) \cdot \alpha_{\w} \quad \text{or, } \quad S_{final}^{DA} = \sum_{\w} \Delta_{\w} \cdot \alpha_{\w},
\end{equation}
where the coefficients $\alpha_\w \in \mathbb{R}$ are chosen according to an aggregation strategy. A final decision based on this aggregation is performed by applying a threshold to $S_{final}$. These two approaches will later be referred to as Aggregate first - Decide next (AD) for the left one and Decide first - Aggregate next (DA) for the one on the right.
In \cite{tutt2024pattern,tutt2022wifs}, the (AD) aggregation was performed according to two simple criterions:

\begin{enumerate}
    \item[(S1)] \emph{Averaging}: Taking the average of all statistics with equal contribution:
    $\alpha_\w = 1.$
    \item[(S2)] \emph{Reliable patterns} \cite{tutt2022wifs}: Selecting only those patterns for which $P_b(\w) \leq \mu$ for a fixed threshold $\mu$ and taking the average of all statistics on these patterns:

    \begin{equation*}
    \alpha_\w = \begin{cases}
    1, &\text{if $P_b(\w) \leq \mu$,}\\
    0, &\text{else.}
    \end{cases}
    \end{equation*}
\end{enumerate}

These approaches, although showing promising results, were chosen empirically and lacked a deeper investigation which could lead to a better understanding of the decision process.
In this paper, we now introduce a systematic approach, based on the average probability of error:

\begin{enumerate}
\setcounter{enumi}{2}
\item[(S3)] \emph{Minimal Probability of error}: Fixing a threshold $\nu~\in~[0,1]$ and only aggregating the patterns which have a probability of classification error smaller than $\nu$:
\begin{equation*}
    \alpha_\w = \begin{cases}
    1, &\text{if $P_{err}(\w) \leq \nu$,}\\
    0, &\text{else.}
    \end{cases}
    \end{equation*}

\item[(S4)] \emph{2C-SVM optimal}: Training a 2-class linear SVM \cite{libsvm_2011} on the statistics and extract the coefficients learned by the model as "optimal" machine learning coefficients $\alpha_\w$. We then sort them according to their absolute value and aggregate the $k$ firsts:
$| \alpha_{\w_1} | \geq | \alpha_{\w_2} | \geq | \alpha_{\w_3} | \geq ...$ .

\end{enumerate}

\section{Experimental results}

\subsection{1x1 Indigo base smartphone dataset}

The dataset\footnote{ The dataset is publicly available and can be found here: \url{https://sipcloud.unige.ch/index.php/s/tYKffnKNRgSwBAN}} of CDP that we use in the following experiment is fully presented in \cite{tutt2024pattern}. It consists of 1440 unique triples of digital template, original printed and fake CDP $(\t, \x, \f)$ enrolled with an \emph{iPhone 12 Pro} during 6 capture runs.
The CDP acquired in this way are then processed using histogram matching to a reference CDP and binarized using Otsu's method.
In the following experiments, the dataset is divided into two randomly chosen train set and test set, which consist of 500 triples. The train set is used to compute reference codebooks of originals $\{ P_b(\w) \}_{\w=0}^{M-1}$ and fakes $\{ Q_b(\w) \}_{\w=0}^{M-1}$.

\subsection{Reliability and authentication performance}

\begin{figure}[t]
\centering
\includegraphics[scale=.59, clip]{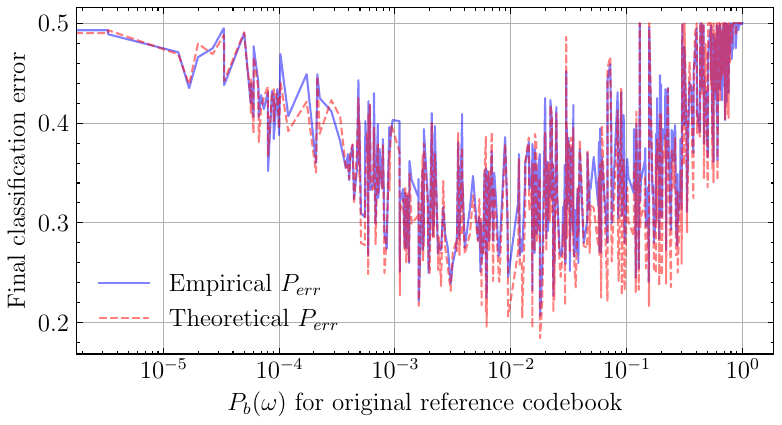}
\caption{Average probability of error between originals and fakes samples as a function of the probability of bit-flipping of each channel. The optimal probability of error is computed using the theoretical model (in red) and empirically on the test set (in blue). Both curves show a close match.}
\label{fig:min_avg_p_error_theory_vs_exp}
\end{figure}

The results of this experiment are shown in Fig. \ref{fig:min_avg_p_error_theory_vs_exp}. For each pattern $\w$ independently, we measure the minimal average classification error $P_{err}$ both theoretical (red dashed curve) by referring to the optimal threshold selection $\gamma_{crit}$ derived in \eqref{eq:def_gamma_crit} for $L^\w = 100$, and experimental (blue curve) by performing an optimal threshold selection on the test set. 

The results show that the theoretical predictions of the average probability of error based on the BPC model closely match the empirical measure based on the testing set. Indeed, the red and blue curve are very close together.

Finally, we also note that pattern reliability, that is low probability of bit-flipping, does not necessarily imply low classification error\footnote{The idea being that an all-white or all-black pattern, while being very reliable ($P_b(\omega)$ is almost 0) is completely useless for classification since fakes will not be wrong when estimating these either and so $Q_b(\omega) = P_b(\omega)$ for those patterns making them indistinguishable from a theoretical point of view.}. Indeed, the plot clearly indicates that the best performing patterns are those located between $P_b(\w) = 10^{-3}$ and $P_b(\w) = 2 \cdot 10^{-2}$. This observation is a clear motivation for us to investigate strategy (S3) based on a bound $\nu$ on $P_{err}$ for the aggregation strategy and compare it with strategy (S2).

\subsection{Aggregation strategies}

\begin{figure}[t]
\centering
\includegraphics[scale=.35, clip]{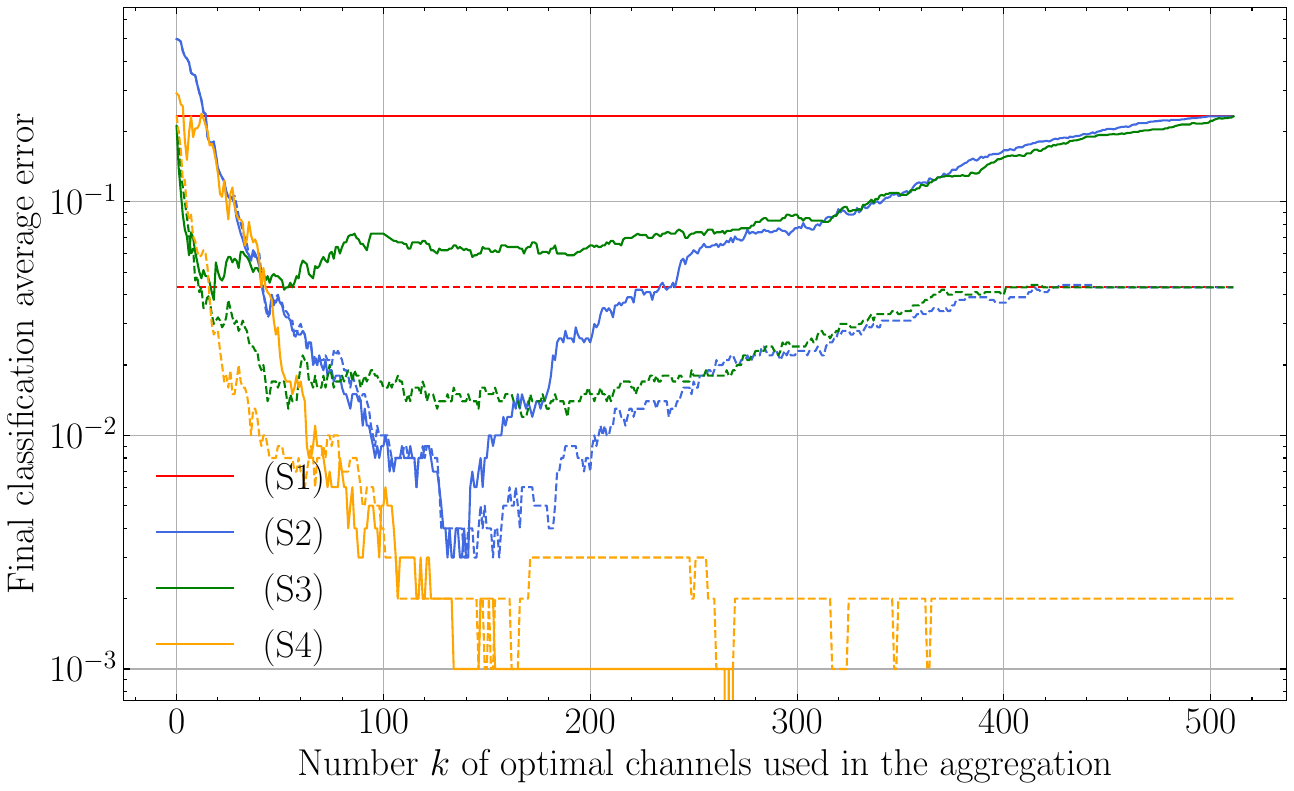}
\caption{A plot of the final classification error probability based on the final score $S_{final}$ for each aggregation strategy as a function of the number of channels aggregated. Continuous curves correspond to (AD) and dashed curves correspond to (DA). The best $k$ channels are different for each strategy and depend on the optimality criterion.}
\label{fig:aggregation_strategy_comparison}
\end{figure}

In this second experiment, we test the various aggregations strategies introduced in Section \ref{sec:aggregation_strategies}. Fig. \ref{fig:aggregation_strategy_comparison} shows the final average error of classification as a function of the number $k$ of patterns used for authentication. The selection of the $k$ best patterns depends on the strategy used for aggregation and is fully described in Section \ref{sec:aggregation_strategies}.

Surprisingly, even though the strategy (S3) based on the minimal $P_{err}$ is optimal for each pattern taken individually, aggregating according to this strategy does not improve the final decision score on this dataset. Indeed, the aggregation strategy (S2) based on the minimization of $P_b(\w)$ introduced in \cite{tutt2024pattern} still outperforms this theoretically optimal criterion, while still being solely based on a reference codebook of originals.
Finally, the fully supervised aggregation scheme (S4) based on linear 2-class SVM performs optimally and, contrary to the other aggregation schemes, does not worsen when the number of patterns used gets too big. This approach, although theoretically interesting, requires the knowledge of both original CDP and fake CDP to be applied and we conjecture that it should be prone to errors against other types of unseen fakes \cite{pulfer2022icip}.

\subsection{Final classification performance}

In this third experiment, we compare the final aggregated average classification error for the different aggregating strategies introduced in Section \ref{sec:aggregation_strategies}. A direct comparison is done between single-shot aggregated performance and multi-shot aggregated performance, where all 6 capture runs are used and fused before computing the Hamming distance test, effectively increasing the number $L^\w$ of symbols in each CDP by a factor of 6. (The behaviour of these symbols is of course strongly correlated but still enhances the prediction capacity.)

When looking at Table \ref{table:perror_strategy_comparison}, a striking observation is the outstanding enhancement of performance when using multi-shot prediction. Indeed, all aggregation strategies benefit from this approach. We analyze this improvement as an illustration of Prop. \ref{thm:likelihood_test_asymptotic}.
Without much surprise, the fully supervised linear 2C-SVM aggregation strategy (S4) performs the best. It is still remarkable to notice that such an approach is able to perfectly separate originals and fakes when all other approaches still miss by a few.
Finally, the (DA) approach also shows an interesting enhancement in performance compared to the (AD) approach which was used in the previous papers.

\section{Conclusion}
\label{sec:conclusion}
In this article, we presented a theoretical analysis of the Binary Pattern-based Channel Model \cite{tutt2024pattern}. We derived optimal bounds for each channel for the classification task and introduced a structured description for the aggregation strategies of the different channels. By comparing with a real dataset of smartphone acquired CDP, we demonstrated that the BPC model correctly simulates the authentication process and allows to reason about the design of optimal aggregation strategies. 
In future works, we plan to pursue the analysis of the interplay between the theoretical BPC model and the real life CDP datasets. Extending this research to new types of fakes and continue to explore the design of optimal aggregation strategies.

\begin{table}[t]
\caption{Results of minimal average $P_{err}$ (in percent) for the aggregation of channels based on four different strategies with Aggregate-Decide (AD) or Decide-Aggregate (DA). Both single-shot and multi-shot settings are compared.}
\center
\begin{tabular}{ccccccc}
\toprule
 &  &  & \multicolumn{4}{c}{Aggregation Strategy} \\
 &  &  & (S1) & (S2) & (S3) & (S4) \\
 \midrule
\multirow{3}{*}{\emph AD} & single-shot &  & $20.29$ & $0.36$ & $3.07$ & $0.10$  \\
\\
 & multi-shot &  & $14.30$ & $0.04$ & $0.13$ & $0.00$ \\
\\
\multirow{3}{*}{\emph DA} & single-shot &  & $4.12$ & $0.40$ & $0.96$ & $0.11$ \\
\\
 & multi-shot &  & $1.20$ & $0.05$ & $0.06$ & $0.00$ \\
\bottomrule
\end{tabular}
\label{table:perror_strategy_comparison}
\end{table}

\bibliographystyle{IEEEtran}
\bibliography{bibliography}

\end{document}